\documentclass[12pt,letterpaper]{article}
\usepackage{graphicx}
\usepackage{hyperref}
\renewcommand{\baselinestretch}{1.4}
\begin{document}
\begin{center}
\bfseries
\large
Quantum Criticality:\\ Competing Ground States in Low Dimensions\\
\normalsize \mdseries \vspace{0.2in}
Subir Sachdev${}^{\dagger}$\\
~\\
Department of Physics, Yale University, P.O. Box 208120, \\New Haven, CT~06520-8120,
USA\\
~
\end{center}
\begin{quote}
${}^{\dagger}$ URL:
\href{http://pantheon.yale.edu/~subir}{http://pantheon.yale.edu/{\~\/}subir};
E-mail: subir.sachdev@yale.edu
~\\
Published in {\sl Science} {\bf 288}, 475 (2000).\\
This cond-mat version has updated references and new footnotes
\cite{new0,new1,new2,new2a,new3}.
\end{quote}
\begin{center}
ABSTRACT
\end{center}
\begin{quote}
Small changes in an external parameter can often lead to dramatic
qualitative changes in the lowest energy quantum mechanical ground
state of a correlated electron system. In anisotropic crystals,
such as the high temperature superconductors where
electron motion occurs primarily on a two-dimensional square lattice,
the quantum critical point between two such lowest energy states has non-trivial
emergent excitations which control the physics over a
significant portion of the phase diagram.
Non-zero temperature dynamic properties
near quantum critical points are described
using simple theoretical models. Possible quantum phases and transitions
in the two-dimensional electron gas on a square lattice are
discussed.
\end{quote}
\newpage

\section{Introduction}
\label{intro}

Quantum mechanics was originally developed by Schr\"{o}dinger and
Heisenberg as a theory of non-relativistic charged particles
interacting via the Coulomb force, and successfully applied to a
simple two-particle system like the hydrogen atom. However, among
its most important applications has been the description of $\sim
10^{23}$ particles found in macroscopic matter. The earliest
example of this was the Sommerfeld-Bloch theory of electronic
motion in metals, and its refined formulation in Landau's Fermi
liquid theory \cite{landau}. Although solving Schr\"{o}dinger's
wave equation for $10^{23}$ interacting electrons appears an
impossibly daunting task, Landau outlined a powerful strategy,
involving the concept of ``quasiparticles'', which allowed an
essentially exact description of the low temperature ($T$)
properties of metals. Extensions of Landau's approach have
successfully described many other phases of matter: the
superfluid phases of $^{4} {\rm He}$ and $^{3} {\rm He}$, the
superconductivity in metals which is described by the
Bardeen-Cooper-Schrieffer theory, and the quantum Hall liquid
state of electrons in two dimensions in a strong magnetic field.
However, in the last decade, attention has been lavished on new
transition metal compounds for which no successful
quasiparticle-like theory has yet emerged for much of the
accessible temperature range. The most important among these
compounds are ceramics, like ${\rm Y Ba}_2 {\rm Cu}_3 {\rm O}_7$,
in which the electronic motion occurs primarily in
two-dimensional ${\rm Cu O}_2$ layers, and which display ``high
temperature'' superconductivity.

In this article we shall describe a new approach to the collective
dynamical properties of electrons which turns out to be especially
useful in two dimensions: the approach focuses on the notion of
competing ground states and its implications for the dynamics of
excited states at non-zero temperatures. Before describing this
further, let us review the essence of Landau's strategy.
His starting point is the proper identification of the quantum
`coherence' or order in the ground state of the system.
In the theory of metals, the order is that implied by the
distribution in the occupation number of plane wave states of
electrons---the plane waves with small wavevectors are fully
occupied, but there is an abrupt decrease in the average
occupation number above a certain ``Fermi wavevector''; in the
superfluid state of $^{4} {\rm He}$, the order in the ground state
is the presence of the Bose-Einstein condensate---the macroscopic
occupation of $^{4} {\rm He}$ atoms in the ground state. Landau
then proceeds to describe the low energy excited states, and hence
the finite temperature properties, by identifying elementary
excitations which perturb the order of the ground state in a
fundamental way. These excitations can be thought of as new,
emergent particles (or `quasiparticles') which
transport spin, charge, momentum, and energy, and whose mutual
collisions are described by a Boltzmann-like transport
equation. In metals, the quasiparticles are electrons and holes in
the vicinity of the Fermi wavevector, while in $^{4} {\rm He}$ they are
phonon and roton excitations.

The systems we shall consider in this article are delicately
poised between two or more distinct states with very different
quantum ordering properties and low-lying excitations. The
energies of the states are quite close to each other, and only at very
low temperatures is a particular one picked as the ground
state---at these temperatures Landau's quasiparticle approach
can apply.
However, for somewhat different parameters, it is possible that a different
state will be picked as ground state, and again Landau's
quasiparticle approach will apply at very low temperatures:
a crucial point is that the nature and
physical properties of these quasiparticles
will, in general, be very different from the
previous ones.
At slightly higher temperatures it is impossible to
ignore the competition between the different states and their respective
quasiparticles: the simple
quasiparticle picture breaks down, and very complex behavior can
result which is not characteristic of any one of the possible
ground states.

We describe this intricate temperature dependence by the following
strategy. Imagine following the true ground state of the system as
a function of some parameter in the Hamiltonian, $g$. It should
be possible to find a critical value $g=g_c$ such that the ground
state undergoes a quantum phase transition \cite{sondhi,book}
from one possible state for $g<g_c$ to another, with distinct
quantum order, for $g>g_c$. We first develop a theory for the
ground state for the quantum critical point precisely at $g=g_c$.
In general, this is a difficult task, but for ``second order''
quantum transitions, the critical point has special symmetry
properties which often allows significant progress---we will see
examples of this below. Empowered with this knowledge of the
physics at intermediate coupling, we move away from the critical
point and  map out the physics for nonzero $|g-g_c|$ and
temperature. It should be emphasized that it is often the case
that the point $g=g_c$ is in a regime that cannot be
experimentally accessed; however, this does not rule out
application of our strategy---it is still useful to describe the
physics at the inaccessible point $g=g_c$, and then use it as a
point of departure to develop a systematic and controlled theory
for an accessible value of $g$.

Our discussion has so far been rather abstract; we will now spell
out concrete details by considering a number of examples of
increasing complexity and discussing their relationship to
experimental observations.

\section{Ising chain in a transverse field}
\label{ising}

This is the simplest theoretical model of a quantum phase
transition and many key concepts emerge from its study
\cite{book}. It is described by the Hamiltonian ($J>0$, $g>0$)
\begin{equation}
H_I = - J \sum_j \left ( g \hat{\sigma}^{x}_{j} +
\hat{\sigma}^{z}_{j} \hat{\sigma}^{z}_{j+1} \right).
\label{hi}
\end{equation}
Here $\hat{\sigma}^{x,z}_j$ are Pauli matrices
which measure the $x,z$ components of the electron spin on a magnetic ion
in an insulator. The ions reside
on the sites $j$ of
a one-dimensional chain. Each site has two possible states $|\uparrow \rangle_j$
and $| \downarrow \rangle_j$, which are eigenstates of
$\hat{\sigma}^{z}_{j}$ with eigenvalues $+1$ and $-1$, and thus
identify the electron spin on site $j$ as `up' or `down'.
The two terms in $H_I$ represent different physical
effects--the second term prefers that the spins on neighboring
ions are parallel to each other, while the first allows
quantum tunneling between the $|\uparrow \rangle_j$ and
$|\downarrow \rangle_j$ states with amplitude proportional to $g$.

\begin{figure}[!ht]
\centerline{\includegraphics[width=5.5in]{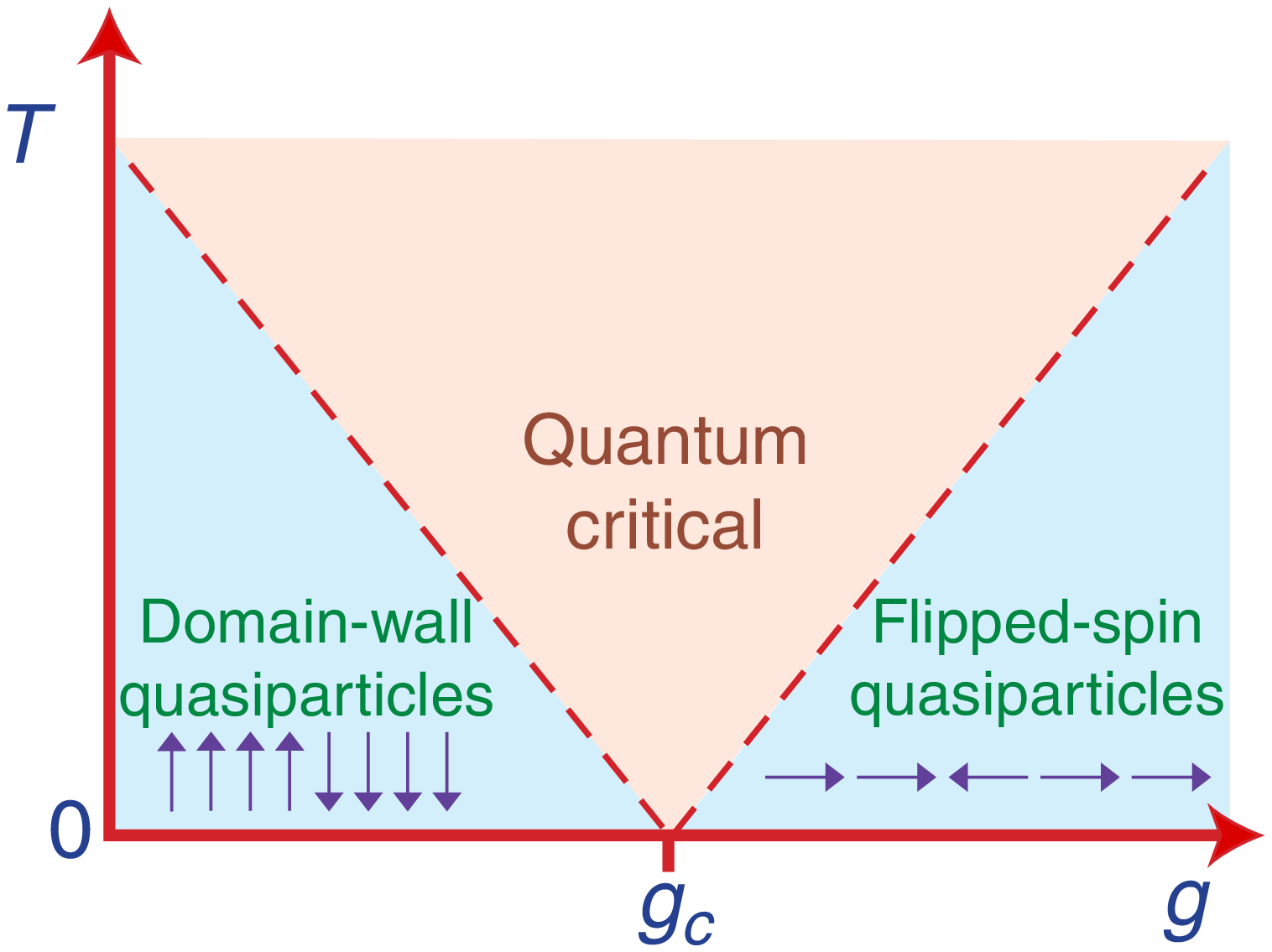}}
\vspace{0.3in} \caption{Phase diagram of $H_I$. The quantum phase
transition is at $g=g_c$, $T=0$, and the dashed red line
indicates a crossover. Quasiparticle dynamics applies in the blue
shaded regions: for $g<g_c$ the quasiparticle states are like the
$|Q_j \rangle$ states, while for $g > g_c$ they are like the very
different $|\widetilde{Q}_j \rangle$ states. The quantum critical
dynamics in the pink shaded region is characterized by
Eq~\protect\ref{scale}. } \label{fig1}
\end{figure}

For $g \gg 1$ and for $g \ll 1$, the ground states of $H_I$ are
simple, and the quasiparticle picture does describe the low $T$
dynamics \cite{apy}. For $g \ll 1$ we can neglect the quantum
tunneling and the ground state either has all spins up or all
spins down. The order in this state is evident--all the spins are
parallel to each other. The quasiparticles are domain walls which
perturb this order---a quasiparticle state, $|Q_j \rangle$,
between sites $j$ and $j+1$ has the following wavefunction: all
spins at and to the left (right) of site $j$ ($j+1$) are up
(down) (see Fig~\ref{fig1}). For $g=0$ every such spin
configuration is an energy eigenstate and therefore stationary;
for small but finite $g$ the domain walls become mobile (and
acquire zero point motion). A theory for the quantum kinetics of
these particles, describing their collisions, lifetime, and the
relaxation of the magnetic order, can be developed following
Landau's general strategy. In the opposite limit, $g \gg 1$, we
see from Eq~\ref{hi} that the ground state can be built out of
eigenstates of $\hat{\sigma}^{x}_i$ with eigenvalue $+1$: these
are
\begin{equation}
|\rightarrow \rangle_j = \frac{1}{\sqrt{2}} \left( |\uparrow
\rangle_j + | \downarrow \rangle_j \right),
\label{right}
\end{equation}
or a `right' pointing spin, which quantum mechanically is just a
linear superposition of up and down spins. The ground state has
all spins pointing to the right, and it is evident that such a
state is very different from the $g=0$ ground state, as the two
states form distinct quantum
superpositions of the available states in the Hilbert space. The
distinction extends also to the excited states:
we define a left pointing spin by the analog of (\ref{right}),
$| \leftarrow \rangle_j = (|\uparrow
\rangle_j - | \downarrow \rangle_j)/\sqrt{2}$, and the
quasiparticle states, $|\widetilde{Q}_j \rangle$,
now represent a single `left'
pointing spin at site $j$ in a background of `right' spins,
rather than a domain wall (see Fig~\ref{fig1}).
For $g=\infty$ these states are stationary, but for
$g < \infty$ the quasiparticles develop dynamics; a theory for this
dynamics can again be formulated
in the spirit of Landau, and this describes relaxation
phenomena at low $T$.

We now allow competition between the distinct orders at small and
large $g$ by considering values of $g$ of order unity. Consider
first $T=0$. It is known that there is a quantum phase transition
between these states at $g=g_c = 1$, {\em i.e.} the ground state
qualitatively similar to the $g=0$ ground state for all $g < 1$,
while a state like the $g=\infty$ ground state is favored for $g
> 1$. The ground state precisely at $g=g_c$ is very special: it
cannot be characterized by any such simple cartoon pictures. Its
fundamental property is one of scale invariance, as is apparent
from the ground state correlation function \cite{pfeuty}
\begin{equation}
\langle \hat{\sigma}^{z}_j \hat{\sigma}^{z}_k \rangle \sim
\frac{1}{|j-k|^{1/4}} \mbox{~~~for large $|j-k|$}:
\label{ons}
\end{equation}
this power-law decay has the property that the functional form of
the correlation is only modified by an overall prefactor if we
stretch the length scale ({\em i.e.} perform a scale transformation)
at which we are observing the spins. In other words, it is not
possible by to tell by an examination of the ground state
wavefunction how far apart any pair of well-separated spins are.
At $T>0$, a new time scale does appear, namely $\hbar/k_B T$, and
a fundamental property of the quantum critical point of $H_I$ is
that this time scale (involving nothing but the temperature and
fundamental constants of nature) universally determines the
relaxation rate for spin fluctuations. This is made more precise by
examining the zero-momentum dynamic
response function
\begin{equation}
\chi (\omega) = \frac{i}{\hbar} \sum_k \int_0^{\infty} d t
\left\langle[ \hat{\sigma}^z_j (t) ,
\hat{\sigma}^z_k (0)] \right\rangle e^{i \omega t} ,
\end{equation}
where $\hat{\sigma}^z_j (t)$ is an operator at time $t$ in the
Heisenberg picture, and $[,]$ represents a quantum commutator.
The arguments above and
simple dimensional considerations following from Eq~\ref{ons} imply that
for low temperatures $\chi$ obeys
\begin{equation}
\chi (\omega) \sim T^{-7/4} \Phi_I ( \hbar \omega / k_B T),
\label{scale}
\end{equation}
with $\Phi_I$ a universal response function; if, for example, we
added a small second neighbor coupling to $H_I$, the critical
coupling $g_c$ would shift slightly but $\Phi$ would remain
exactly the same. The exact result for $\Phi_I$ is known, and it
is an excellent approximation to just replace its inverse by a
low frequency expansion \cite{book} $\Phi_I ( \hbar \omega / k_B
T) = A ( 1 - i \omega / \Gamma_R + \ldots)^{-1}$; here $A$ is a
dimensionless prefactor, and we have the important result that $
\Gamma_R = ( 2 \tan (\pi/16)) k_B T /\hbar$. This is the response
of an overdamped oscillator with a relaxation rate determined
only by temperature itself \cite{SY}. Although a quasiparticle
description of this response function is strictly not possible,
we can visualize the dynamics in terms of a dense gas of the
$|Q_j \rangle$ particles scattering off each other at a rate of
order $k_B T/\hbar$; however, a picture in terms of the `dual'
$|\widetilde{Q}_j \rangle$ particles would also be valid. It is
quite remarkable that the strength of the underlying exchange
interaction between the spins does not appear in these
fundamental dynamic scales.

We can use these above results to sketch a crossover phase diagram
in the $g,T$ plane. This is shown in Fig~\ref{fig1}. Note that
``quantum criticality'', characterized by responses like
Eq~\ref{scale} holds over a range of values of $g$ at non-zero
temperature \cite{CHN}.

It can be shown that the physics of the quantum Ising model in
spatial dimension $d=2$ is very similar--quantum criticality is
again characterized by Eq~\ref{scale} (but the exponent $7/4$ is
replaced by a different universal numerical value). Similar
behavior applies also to quantum transitions in $d=3$ systems with
quenched disorder~\cite{gruner}. However, for the analog of $H_I$
in $d=3$ (and for all $d>3$), the physics of the quantum phase
transition is very different \cite{book}--- the kinetic theory of
the analog of the $|\widetilde{Q}_j \rangle$ quasiparticles
applies even at the critical point, and their scattering
cross-section depends on the magnitude of the microscopic
interactions. Quantum transitions in this class have been studied
elsewhere, and have important physical applications
\cite{hertz,millis,hyper}: our discussion of quantum criticality
will not apply to them.

\section{Coupled ladder antiferromagnet}
\label{ladder}

\begin{figure}[!ht]
\centerline{\includegraphics[width=5.5in]{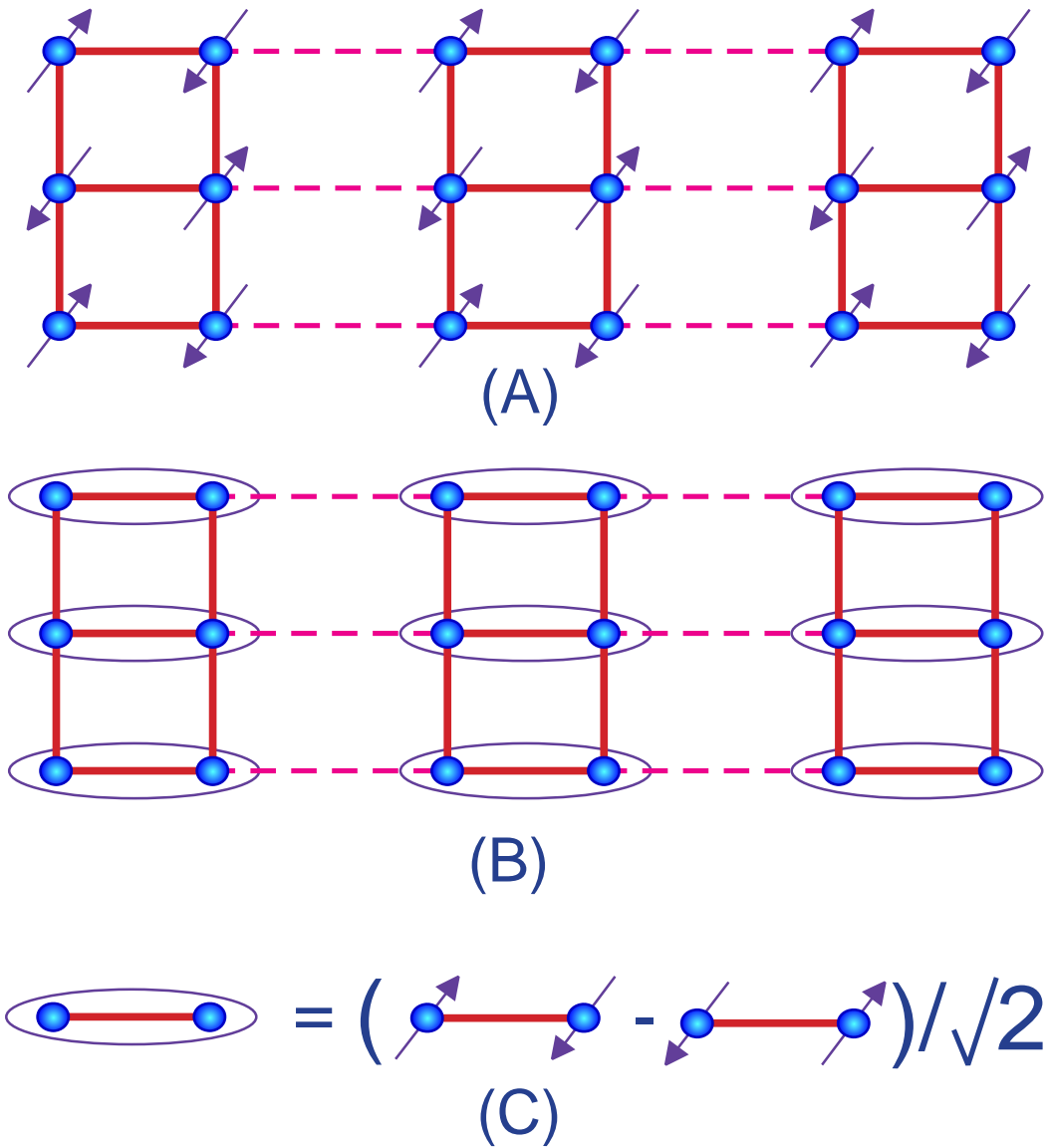}}
\vspace{0.3in} \caption{The coupled ladder antiferromagnet--the
spin-1/2 degrees of freedom, ${\bf S}_i$, reside on the blue
circles. The $A$ links are the full red lines and have exchange
$J$, while the $B$ links are dashed lines and have exchange $g
J$.  The N\'{e}el ground state for $g > g_c$ appears in (A). The
paramagnetic ground state for $g < g_c$ is schematically
indicated in (B). The ellipses in (B) represents a singlet
valence bond, $(|\uparrow \downarrow \rangle - |\downarrow
\uparrow \rangle)/\protect\sqrt{2}$ (shown in (C)), between the
spins on the sites.} \label{fig2}
\end{figure}

We turn to a model in $d=2$ which is indirectly related to
microscopic models of the high temperature superconductors. We
consider the antiferromagnet described by the Hamiltonian
\cite{katoh} ($J>0$, $0 < g \leq 1$)
\begin{equation}
H_L = J \sum_{i,j \in A} {\bf S}_i \cdot {\bf S}_j + g
J \sum_{i,j \in B} {\bf S}_i \cdot {\bf S}_j
\label{ham}
\end{equation}
where ${\bf S}_i$ are spin-1/2 operators on the sites of the
coupled-ladder lattice shown in Fig~\ref{fig2}, with the $A$ links
forming ``two-leg ladders'' while the $B$ links couple the ladders.
There is a quantum phase transition in $H_L$ at a critical value
$g = g_c \approx 0.3$ which is similar in many respects to that in
$H_I$.

We begin by describing the well-ordered ground states on either
side of $g_c$ and their respective, low $T$ quasiparticle
theories.

For $g$ close to unity we have the magnetically ordered
``N\'{e}el'' state in Fig~\ref{fig2}A. This is analogous to the
ground state of $H_I$ for small $g$, with the difference that the
mean moment on the sites has a staggered sublattice arrangement.
There is also an important difference in the structure of the
excitations because $H_L$ has the symmetry of arbitrary rotations
in spin space, in contrast to the discrete spin inversion
symmetry of $H_I$. Consequently the low-lying quasiparticle
excitations are spin waves corresponding to a slow precession in
the orientation of the staggered magnetic order. The precession
can be either clockwise or anti-clockwise, and so there is
two-fold degeneracy to each spin-wave mode. Because of infrared
singular scattering of thermally excited spin waves in $d=2$, the
theory of spin-wave dynamics has some subtleties \cite{CHN,tyc};
nevertheless, the results remain within the spirit of the
quasiparticle picture.

For small $g$, the ground state is a quantum paramagnet, and a
caricature is sketched in Fig~\ref{fig2}B. The average moment on
each site has been completely quenched by the formation of
singlet bonds between neighboring spins. This state is similar in
many respects to the large $g$ ground state for $H_I$. It
requires a finite energy, $\Delta$, to create quasiparticle
excitations by locally disrupting the singlet order (the analog
of flipping a spin for $H_I$): the singlet bond between a pair of
spins can be replaced by a triplet of total spin $S=1$ states,
and the motion of this broken bond corresponds to a three-fold
degenerate quasiparticle state (to be contrasted with the
two-fold degenerate spin wave above). A conventional quantum
Boltzmann equation can be used to describe the low temperature
dynamics of these triplet quasiparticles \cite{kedar,book}.

We sketch the crossover phase diagram in the $g$, $T$ plane in
Fig~\ref{fig3} \cite{CHN} following Fig~\ref{fig1}. For $g \leq
g_c$, quantum criticality appears for $\Delta \ll k_B T \ll J$.
Here, dynamic spin response functions have a structure very
similar to that described near Eq~\ref{scale}: the relaxation
rate $\Gamma_R$ continues to be proportional to $k_B T/\hbar$,
but now only approximate results for the proportionality constant
are available \cite{CSY,book}. If we describe the dynamics in the
basis of the triplet quasiparticles, then these results imply
that the scattering cross-section is universally determined by
the energy $k_B T$ alone \cite{kedar}. As we lower $k_B T$ across
$\Delta$ (for $g < g_c$), this scattering cross-section evolves
as a function of the dimensionless ratio $\Delta/k_B T$ alone,
and for very low $T$ is determined by $\Delta$ alone. One
remarkable consequence of this universal cross-section is that
transport coefficients, like the spin conductance $\sigma_s$
(which determines the spin current produced by the gradient in an
applied magnetic field), are determined by fundamental constants
of nature \cite{fgg} and the ratio $\Delta/k_B T$:
\begin{equation}
\sigma_s = \frac{(g \mu_B )^2}{h} \Phi_{\sigma} \left(
\frac{\Delta}{ k_B T}\right).
\label{cond}
\end{equation}
Here $g$ is the gyromagnetic ratio of the ions carrying the spin,
$\mu_B$ is the Bohr magneton, and $\Phi_{\sigma}$ is a universal
function with no arbitrariness in either its overall scale or in
that of its
argument. Note that, well into the quantum critical region,
$\sigma_s$ is proportional to the universal number $\Phi_{\sigma} (0)$, and so is
determined by constants of nature alone.

Although it is certainly not appropriate to take $H_L$ as a
literal model for the high temperature superconductors, it is
notable \cite{lt22} that many measurements of spin fluctuations
in the last decade display crossovers which are very similar to
those found in the vicinity of the quantum critical point in
Fig~\ref{fig3}. We take this as evidence that the high
temperature superconductors are near a quantum critical point
whose spin sector has universal properties closely related to that
of $H_L$ \cite{CS}: a specific microscopic calculation, involving
competition between the states to be discussed in
Section~\ref{2deg}, which realizes such a scenario was presented
in \cite{vprl}. The evidence has appeared in the following
experiments: ({\em i\/}) the dynamic spin structure factor
measured in neutron scattering experiments \cite{aeppli} at
moderate temperature obeys scaling forms similar to
Eq~\ref{scale}; ({\em ii\/}) as we discuss in Fig~\ref{fig4},
crossovers in the nuclear spin relaxation rate
\cite{imai1,imai2,fujiyama} as a function of carrier density and
temperature match very well with the spin dynamics of the
different regimes in the $g$, $T$ plane in Fig~\ref{fig3}; ({\em
iii\/}) low temperature neutron scattering measurements
\cite{keimer} at higher carrier density show a resolution-limited
peak above a finite energy gap---this is a signal of the
long-lived triplet quasiparticles, like those found at low $T$
for $g<g_c$ in $H_L$; such a peak was argued early on \cite{CSY}
to be a generic property of the vicinity of a quantum critical
point, like that in $H_L$, proposed for the high temperature
superconductors \cite{CS}. A further test of quantum criticality
in the spin fluctuations could be provided by measurements of the
spin conductance and comparison with Eq~\ref{cond}, but such
experiments have not been feasible so far.

\section{Electronic ground states in two dimensions}
\label{2deg}

We have so far discussed simple models of quantum phase
transitions whose physics is now well understood. Here we turn to
more realistic models of the high temperature superconductors. As
we mentioned in Section~\ref{intro}, electronic motion in these
materials occurs primarily in two-dimensional ${\rm Cu O}_2$
layers. The Cu ions are located on the vertices of a square
lattice, and it is widely believed that only the dynamics on a
single 3d Cu orbital is relevant, with the occupation numbers of the
other orbitals being inert. So we are led to consider a
simple tight-binding model of electrons with a single orbital on every
site of a square lattice, along with Coulomb interactions between the
electrons. If the electron density is precisely unity per site,
then the ground state is known to be an insulator with N\'{e}el
order (this corresponds to the state in Fig~\ref{fig2}A at $g=1$)
for the range of parameters
found in the stoichiometric compound
${\rm La}_2 {\rm Cu O}_4$. It is possible to vary the electron
density in the square lattice by doping such a compound to ${\rm
La}_{2-x} {\rm Sr}_x {\rm Cu O}_4$, and then $x$ measures the
density of holes relative to the insulating state with one electron
per site. High temperature
superconductivity is found for $x$ greater than about $0.05$.

Much theoretical work in the last decade has addressed the
physics of this square lattice model for small $x$. We will discuss
various proposals for ground states, with an emphasis on
finding sharp distinctions between them {\em i.e.} distinguishing
states which cannot be smoothly connected by the variation of a
parameter in the Hamiltonian, and which must be separated by a
quantum phase transition. Often, the theoretical debate has been
about different approximation schemes to computing properties of
states which are ultimately equivalent; we avoid such issues
here---indeed, we advocate that a sound approach is to
use a theory for quantum critical points, separating
distinct ground states, to develop a controlled expansion at
intermediate coupling.

A minimal approach to identifying possible ground states is to
assume that they are fully characterized by broken symmetries of
the underlying Hamiltonian {\em i.e.} a simple electron mean-field
theory of the broken symmetry properly identifies the
elementary excitations (however, as discussed above,
this does not rule out highly non-trivial
quantum critical points whose excitations control the physics over
a wide region of the phase diagram). The symmetries which leave
the Hamiltonian invariant (and so may be broken by the ground state)
are time-reversal, the group of spin rotations, the space
group of the square
lattice, and the electromagnetic gauge symmetry related to charge
conservation.
Even in this limited framework, the possibilities are
remarkably rich, and it is entirely possible that they will
provide an explanation for all the experiments. More exotic
ground states have also been proposed, and we will note them
briefly in a subsection below.

One important state has already made an appearance in our
discussion in Section~\ref{ladder}, and is known to be the ground
state $x=0$: the N\'{e}el state sketched in Fig~\ref{fig2}A. It
is apparent by a glance at the staggered arrangement of spins in
Fig~\ref{fig2}A that we can view this state as a density wave of
spin polarization at the wavevector ${\bf K} = (\pi/a, \pi/a)$,
where $a$ is the square lattice spacing. For small $x\neq0$, spin
density waves with a period incommensurate with the underlying
lattice have been observed \cite{incomm}: these states have a
mean spin polarization at a wavevector ${\bf K}$ which varies
continuously away from $(\pi/a,\pi/a)$.

The other ground state of central importance is, of course, the
superconducting state. This is formed by Bose condensation of
electrons in Cooper pairs, which leads to the breaking of the
elctromagnetic gauge symmetry. It is known that the pair
wavefunction has the symmetry of the $d_{x^2-y^2}$ orbital in the
relative co-ordinate of the two electrons. Recently, interest has
focused on the question of whether the pair wavefunction is on the
verge of acquiring an additional imaginary component with
$d_{xy}$ (or possibly $s$) symmetry: such an instability would
also break time-reversal symmetry
\cite{kotliar,sigrist,kirtley,did}. It has been argued \cite{did}
that the quantum phase transition between two such
superconductors could very naturally explain the quantum
criticality, similar to the scaling form (\ref{scale}), observed
in recent photoemission experiments \cite{valla}.

A state which makes a frequent appearance in theoretical studies
is one with ``Peierls'' order \cite{new0,new1}. In models with
half-integral spin per unit cell, such order was argued
\cite{rs1} to be a generic property of any state reached by a
continuous quantum transition with restores the broken spin
rotation symmetry of a N\'{e}el state. The Peierls order is
associated with broken translational symmetries, and examples are
shown in Fig~\ref{fig5}: in these states all sites of the square
lattice are equivalent, but links connecting nearest neighbor
sites spontaneously can acquire distinct values for their charge
and energy densities (and therefore, for the mean value of the
exchange coupling $\langle {\bf S}_i \cdot {\bf S}_j \rangle$).
We also considered a quantum transition restoring spin rotation
symmetry in Section~\ref{ladder}, and mentioned its relevance to
the NMR measurements in Fig~\ref{fig4}; however, the issue of
spontaneous Peierls ordering did not arise there because the
links were already explicitly inequivalent in the Hamiltonian
$H_L$ in Eq~\ref{ham}. It is believed \cite{rs1,CSY,ko} that the
universal spin fluctuation properties in the vicinity of the
quantum critical point of Section~\ref{ladder} apply also to
cases where spontaneous Peierls order appears in the paramagnetic
state. Evidence for the spontaneous Peierls ordering in
Fig~\ref{fig5}A has emerged in numerical studies
\cite{rajiv,singh} of square lattice models at $x=0$ and with
first and second neighbor hopping of electrons, and also in
nearest neighbor hopping models for $x>0$ \cite{ops}.

A competitor state to Peierls order for the quantum paramagnet is
the ``orbital antiferromagnet'' \cite{sf,schulz,nayak}: this
state breaks time-reversal and translational symmetries, but spin
rotation symmetry and the combination of time-reversal and
translation by an odd number of lattice spacings remains
unbroken. There is a spontaneous flow of electrical currents
around each plaquette of the square lattice, with clockwise and
anti-clockwise flows alternating in a checkerboard pattern--see
Fig~\ref{fig6}. Ivanov {\em et al.} \cite{wenlee} have proposed
that a closely related state (in their formulation, there are
strong fluctuations of the orbital currents, but no true
long-range order) is responsible for the ``pseudo-gap''
phenomenology of the high temperature superconductors---the
pseudo-gap is the partial quenching of low energy spin and
fermionic excitations at temperatures above the superconducting
critical temperature.

For our final conventional state, we consider a charge density
wave. Much recent experimental work has centered around the
discovery of charge ordering in certain high temperature
superconductors and related materials \cite{tran}. An especially
stable state, observed for $x \approx 1/8$, has a charge density
wave at wavevector ${\bf K} = (\pi/(2a),0)$; depending upon its
phase,
 the charge density wave can be either
site-centered or bond-centered, as shown in Fig~\ref{fig7}.
Current experiments do not distinguish between these two
possibilities. Site-centered ordering was considered in some
early theoretical work \cite{jan,schulz2,machida}, although with
a very different charge distribution than is now observed.
Bond-centered ordering has also been considered
\cite{emery,white,vprl,new0}, and has some attractive features: it
enhances singlet-bond formation between spins, optimizing the
energy gained through quantum fluctuations in an antiferromagnet,
and so is preferred by the same effects that led to the Peierls
ordering in Fig~\ref{fig5}. Also, bond-centering is naturally
compatible with the observed coexistence of charge-ordering and
superconductivity at lower temperatures \cite{trans}, while
site-centering is expected to lead to insulating behavior. Let us
also mention that superposition of charge density waves with
different non-collinear ${\bf K}$ can lead to an insulating
Wigner crystal state; this could be a Wigner crystal of holes, or
with an even number of particles per unit cell, a Wigner crystal
of Cooper pairs \cite{emery}.

Clearly, a fascinating variety of phase diagrams and quantum phase
transitions are possible among the states we have discussed
above. In principle, many of the order parameters can co-exist with each other,
and this adds to the zoo of possibilities. Future experiments
with increased sensitivity should make it possible to more
clearly detect more of these orderings, and thus select between
various scenarios.

\subsection*{Spin-charge separation}
Finally, we discuss more exotic possibilities of states which
cannot be completely characterized by ordering discussed above.
Of particular interest has been the early proposal of Anderson
and others \cite{ab,krs} that there could be an insulating state
with spin-charge separation {\em i.e.} the electron falls apart
(`fractionalizes') into separate deconfined excitations
\cite{rs2,wen} which carry its spin and charge. A fundamental
property of these deconfined phases is that superconducting
states in their vicinity allow low energy vortex excitations with
quantized magnetic flux equal to $hc/e$
\cite{vortex,nodal,senthil,new2}: the elementary flux quantum is
always $hc/2e$, but it can be argued quite generally that core
energy of a $hc/e$ vortex is lower than twice that of a $hc/2e$
vortex. The kinetic energy of the superflow well away from the
vortex core always prefers the smaller flux $hc/2e$, and so
requiring global stability for $hc/e$ vortices becomes a delicate
question of balancing core and superflow contributions
\cite{vortex,new2}. Nevertheless, it is possible in principle that
a magnetic flux decoration, flux noise, tunneling, or other
experiment could observe metastable or stable $hc/e$ vortices or
vortex-anti-vortex pairs: this would be a ``smoking-gun'' signal
for deconfinement \cite{new2a}.

Another important class of experiments \cite{alloul} measure the
response to non-magnetic impurities like Zn or Li, and these
could also provide clear-cut answers on the nature of the order
parameter and on issues of confinement. These impurities
substitute on the Cu site, and so are directly within the plane
of the two-dimensional electron gas. Such deformations are very
effective in disrupting the quantum coherence of the ground
state, and so serve as effective probes of its structure. For
example \cite{fong,qimp}, replacing only 0.5\% of Cu by Zn
dramatically broadens the peak in the neutron scattering
cross-section \cite{keimer}, arising from the triplet
quasiparticles discussed in Section~\ref{ladder}. Further
interesting developments are sure to follow \cite{new3}.

\section{Conclusion}
\label{conc} Correlated electron systems in two dimensions are in
a privileged position. Those in three dimensions either form good
Fermi liquids or ordered states with order parameters like those
in Section~\ref{2deg}: in the latter case, quantum fluctuations
of the order parameter are weak and do not lead any unusual
non-quasiparticle behavior even at zero temperature phase
transitions \cite{hyper}. In contrast, in one dimension, quantum
fluctuations of the order parameters are so strong that they
usually preclude the emergence of long-range order, and so
quantum phase transitions are harder to find. It is in two
dimensions that there is a delicate balance between order and
fluctuation, and a host of interesting quantum critical points,
with non-trivial universal properties, can appear between
different competing orders. This article has considered some
simple examples of the dynamical properties of systems near such
a point. The phases on either side of the critical point are
usually amenable to a quasi-particle description at low enough
temperatures. However, a key point is that the quasiparticle
states are very different for the two phases: so at slightly
higher temperatures when both phases can be thermally excited,
neither quasiparticle description is appropriate. Instead special
scale-invariance properties of the critical point have to be used
to develop a new framework for finite temperature dynamics.

The availability of a large number of two-dimensional correlated
electron systems (including the high temperature superconductors),
along with the highly non-trivial theoretical framework necessary
to describe them, makes this one of the most exciting research
areas in condensed matter physics. As we have already noted,
the increased sensitivity of future experiments, including neutron
scattering, tunneling , magnetic resonance, photoemission and optics, along
with better sample preparation techniques, will surely uncover
much new physics. Many interesting theoretical questions, on the classification
of ground states and quantum critical points, and on the
description of dynamical crossovers in their
vicinity, remain
open. The interplay between theory and experiment promises to
be mutually beneficial, in the best traditions of physics
research.

\newpage
\renewcommand{\baselinestretch}{1.2}
\section*{Figures}

\begin{figure}[!ht]
\centerline{\includegraphics[width=5.5 in]{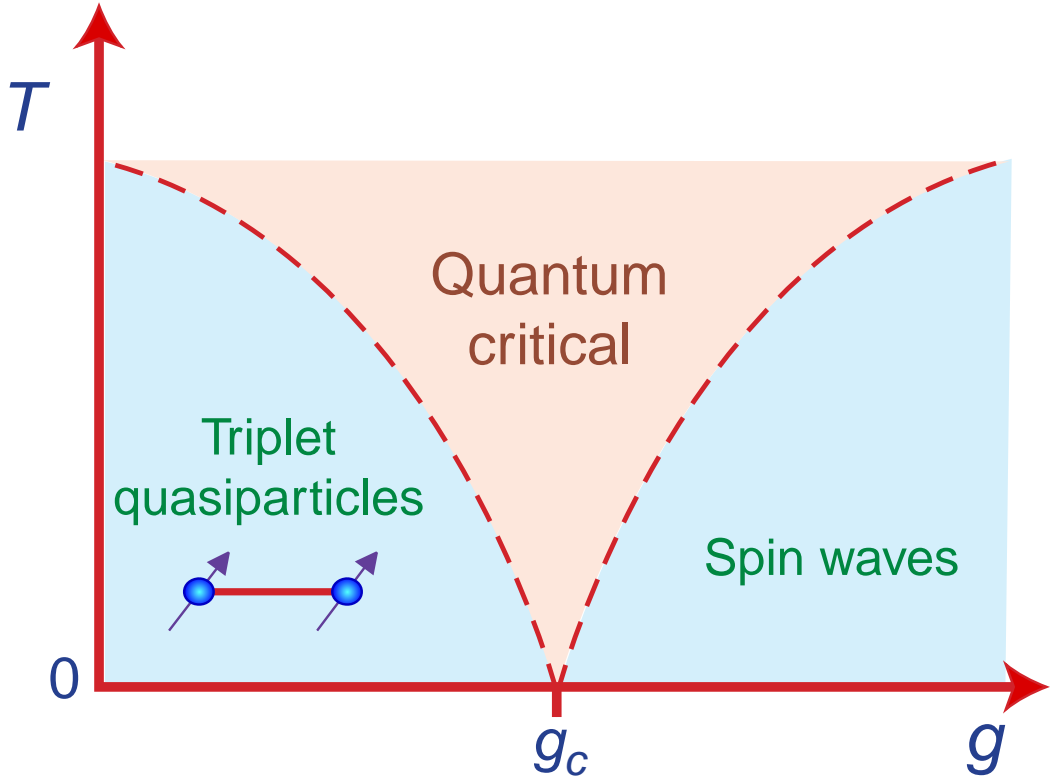}}
\vspace{0.2in} \caption{Crossover phase diagram
\protect\cite{CHN} for $H_L$ with the same conventions as
Fig~\protect\ref{fig1}. The ground state is a paramagnet
(Fig.~\protect\ref{fig2}B) for $g < g_c$ and the energy cost to
create a spin excitation, $\Delta$ is finite for $g<g_c$ and
vanishes as $\Delta \sim (g_c - g)^{z \nu}$ where $z \nu$ is a
critical exponent. There is magnetic N\'{e}el order at $T=0$ for
$g>g_c$ (Fig.~\protect\ref{fig2}A) and the time-averaged moment
on any site, $\vec{N}_0$, vanishes as $g$ approaches $g_c$ from
above. Quasiparticle-like dynamics applies in the blue shaded
regions. For $g<g_c$, in the cartoon picture of the ground state
in Fig~\protect\ref{fig2}B, the triplet quasiparticle corresponds
to the motion of broken singlet bond in which Fig~\protect{fig2}C
is replaced by one of $|\uparrow \uparrow \rangle$,
$|\downarrow,\downarrow\rangle$, or $(|\uparrow \downarrow
\rangle + |\downarrow \uparrow \rangle)/\sqrt{2}$. For $g > g_c$,
the quasiparticles are spin-waves representing slow,
long-wavelength deformations of the ordered state in
Fig~\protect\ref{fig2}A. } \label{fig3}
\end{figure}

\begin{figure}[!ht]
\centerline{\includegraphics[width=6in]{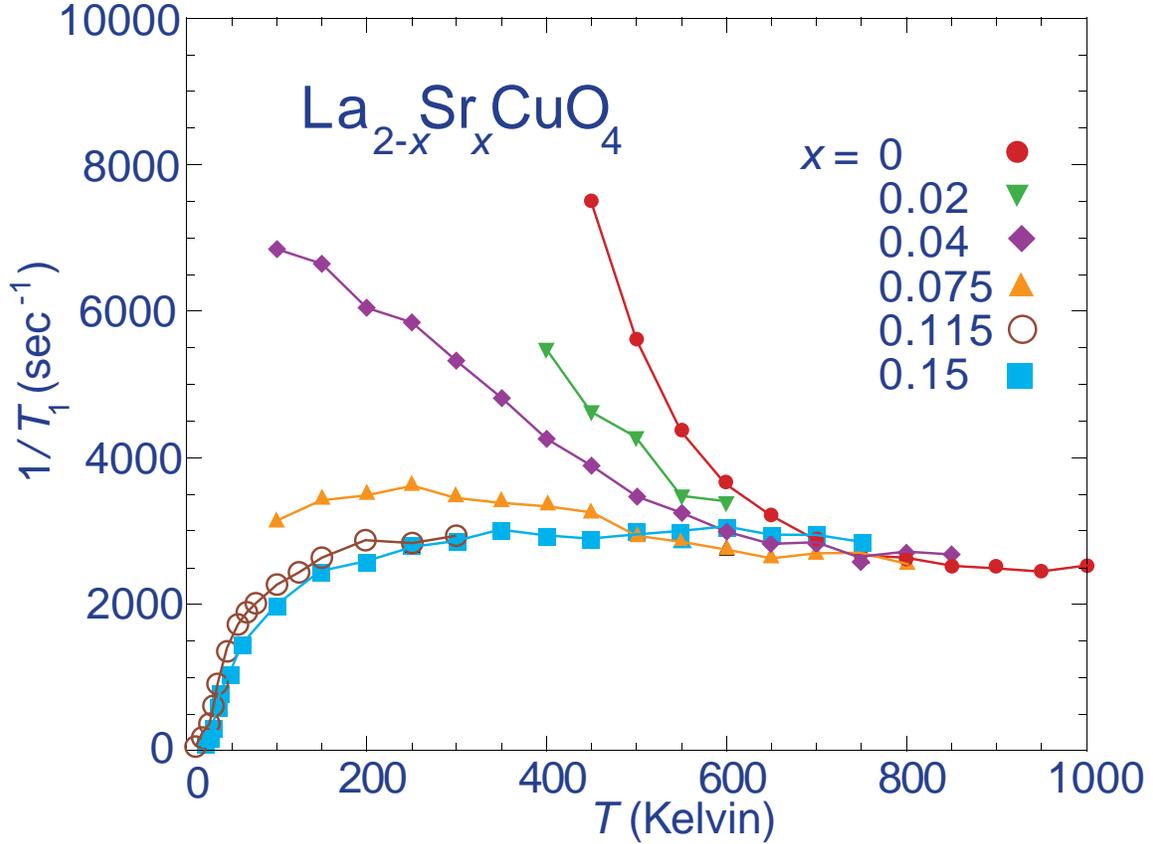}} \vspace{0.2in}
\caption{Measurements \protect\cite{imai1} of the longitudinal
nuclear spin relaxation ($1/T_1$) of $^{63}Cu$ nuclei in the high
temperature superconductor ${\rm La}_{2-x} {\rm Sr}_x {\rm Cu
O}_4$ as a function of $x$ and $T$. This quantity is a measure of
the spectral density of electron spin fluctuations at very low
energies. At small $x$, $1/T_1$ increases rapidly as $T$ is
lowered (see red circles): this is also the behavior in the
spin-wave regime of Fig~\protect\ref{fig3} ($g>g_c$)---the energy
of the dominant thermally excited spin-wave decreases rapidly as
$T$ decreases, and so the spin spectral density rises
\protect\cite{co}. In contrast, at large $x$, $1/T_1$ decreases
as $T$ is lowered (see blue squares): this corresponds with the
triplet quasiparticle regime of Fig~\protect\ref{fig3}
($g<g_c$)---the low energy spectral density is proportionally to
the density of thermally excited quasiparticles, and this becomes
exponentially small as $T$ is lowered. Finally, at intermediate
$T$, $1/T_1$ is roughly temperature independent for a wide range
of $T$ (see orange triangles) and this is the predicted behavior
\protect\cite{CS,sokol} in the quantum critical regime of
Fig~\protect\ref{fig3}. } \label{fig4}
\end{figure}

\begin{figure}[!ht]
\centerline{\includegraphics[width=4in]{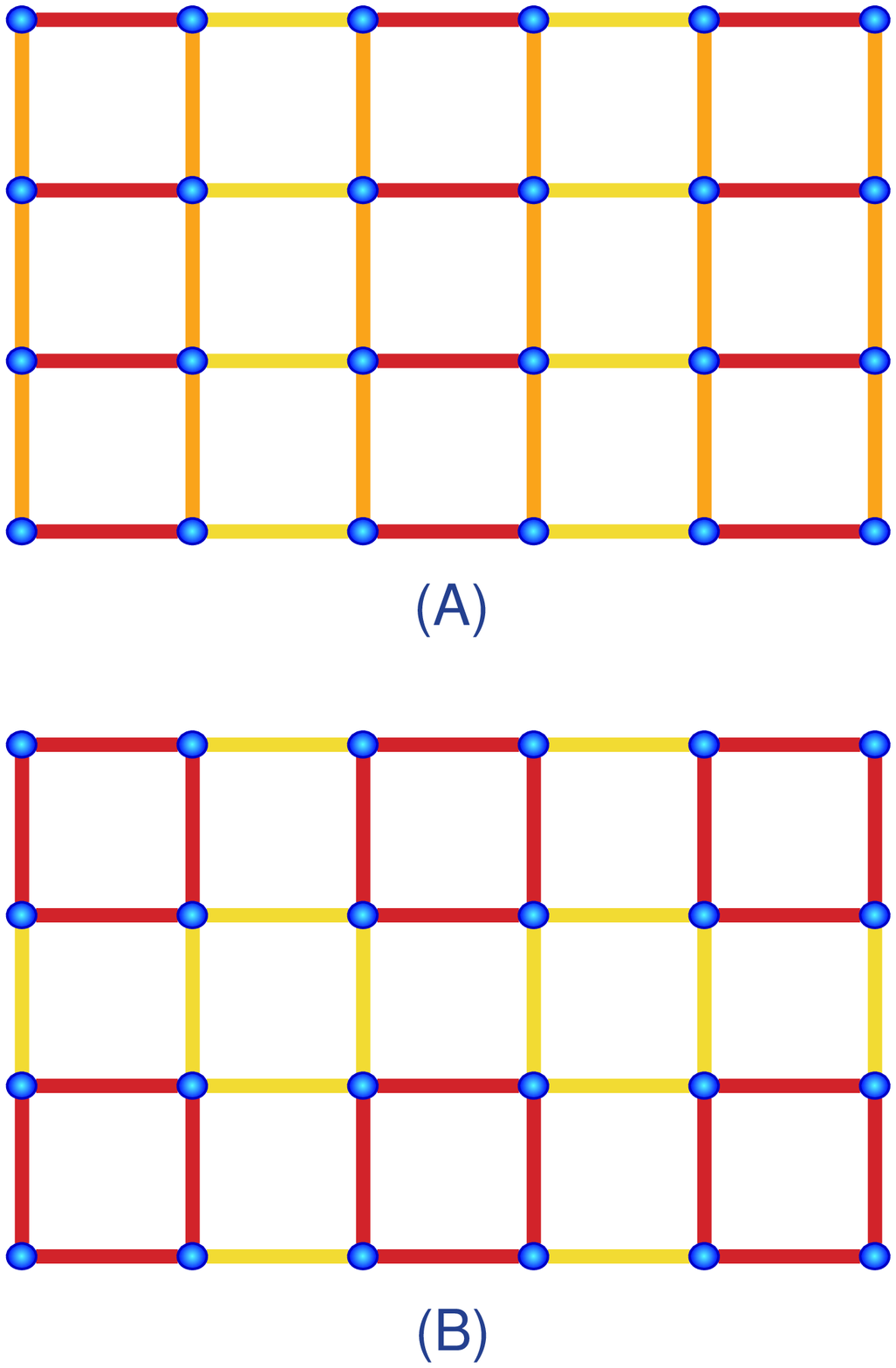}} \vspace{0.2in}
\caption{ Two examples of square lattice ground state with
Peierls order. All sites are equivalent, and distinct values of
the energy and charge densities on the links are represented by
distinct colors. These distinctions represent a spontaneous
breaking of the symmetry of the square lattice space group. The
spontaneous ordering appears because it optimizes the energy
gained by resonance between different singlet bond pairings of
near neighbor spins. This figure should be contrasted with
Fig~\protect\ref{fig2} where there is no spontaneous breaking of
translational symmetry, and the distinction between the links is
already present in the Hamiltonian Eq~\protect\ref{ham}. }
\label{fig5}
\end{figure}

\begin{figure}[!ht]
\centerline{\includegraphics[width=4.0in]{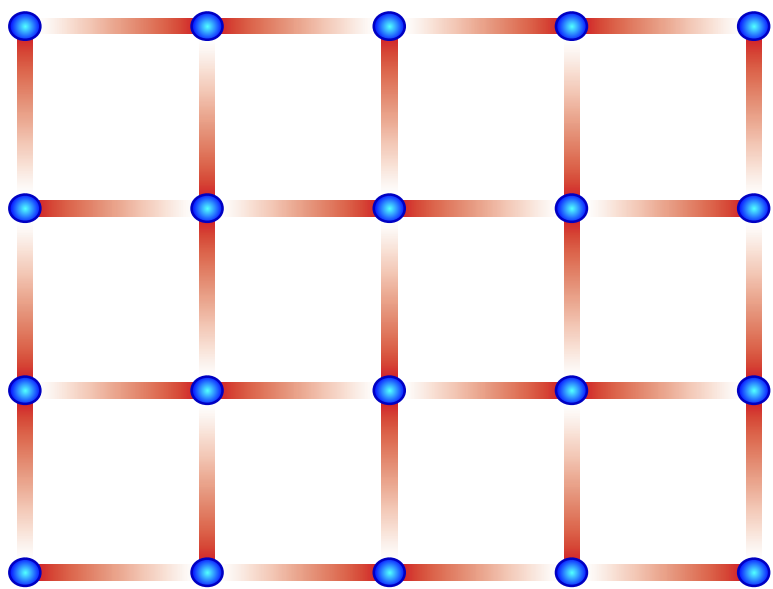}}
\vspace{0.2in} \caption{ The orbital antiferromagent. The
gradient in the red shading represents the direction of
spontaneous current flow on the links, which breaks time-reversal
symmetry. } \label{fig6}
\end{figure}

\begin{figure}[!ht]
\centerline{\includegraphics[width=4.0in]{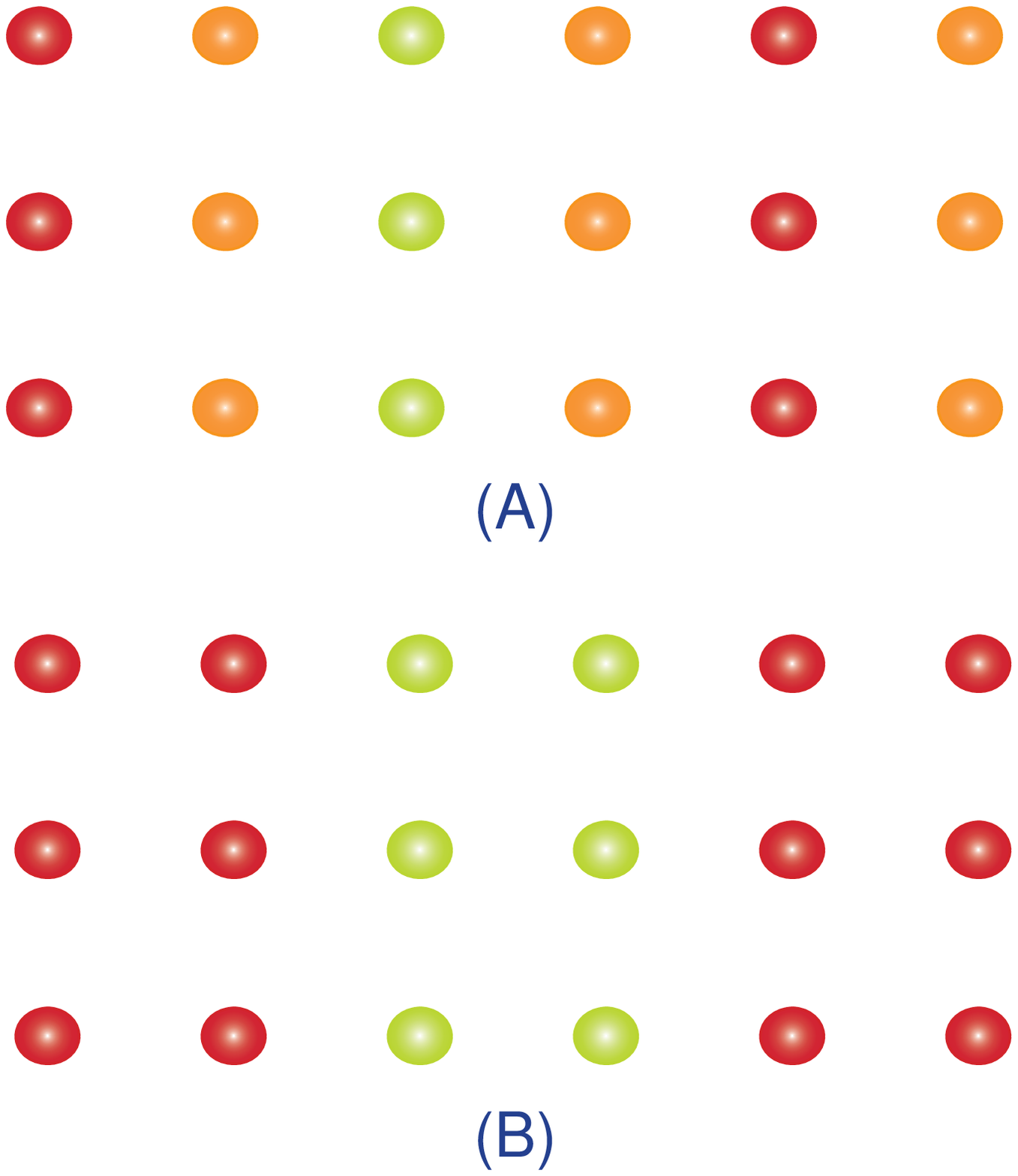}}
\vspace{0.2in} \caption{ Charge density waves with (A) site- and
(B) bond-centering; both states have a $4 \times 1$ unit cell. The
state in (A) is symmetric about reflections in a vertical axis
running through the red or green sites, while that in (B)
requires a vertical axis centered on the bond between two red
sites or two green sites. The colors of the sites represent
different charge densities. Spin ordering can also be present for
an appropriate ${\bf K}$ but is not shown. Note that the
bond-centered ordering naturally suggests an effective model for
the spin fluctuations much like the ladder model of
Section~\protect\ref{ladder}: the spins of Fig~\protect\ref{fig2}
reside on the red sites of (B) (with no spins on the green sites)
and the weaker $g J$ exchange interactions (represented by the
dashed lines Fig~\protect\ref{fig2}) extend across the green
sites. } \label{fig7}
\end{figure}





\begin{thebibliography}{}
\bibitem{hyper} More technically, the quantum criticality
described in the present paper is associated with critical points
which obey hyperscaling and are below their upper critical
dimension; for models with a bosonic vector order parameter, this
is the case when $d+z<4$ where $z$ is the dynamic critical
exponent ($z=1$ for $H_I$).
\end{thebibliography}
\end{document}